# Total controllability analysis discovers explainable drugs for Covid-19 treatment

*Xinru Wei*[1,2,+], *Chunyu Pan*[3,+], *Xizhe Zhang*[1,2,*], *Weixiong Zhang*[4,5,*]

1. Early Intervention Unit, Department of Psychiatry, The Affiliated Brain Hospital of Nanjing Medical University, Nanjing, Jiangsu 210029, China;

2. School of Biomedical Engineering and Informatics, Nanjing Medical University, Nanjing, Jiangsu 210001, China

3. School of Computer Science and Engineering, Northeastern University, Shenyang, China

4. Department of Health Technology and Informatics, Department of Computing, The Hong Kong Polytechnic University, Hong Kong, China

5. Department of Computer Science and Engineering, Department of Genetics, Washington University in St. Louis, St. Louis, MO, USA

+: Equal contribution.

*: Correspondence: XZ: zhangxizhe@njmu.edu.cn, WZ: weixiong.zhang@polyu.edu.hk

## ABSTRACT

Network medicine has been actively pursued for drug repurposing, particularly lately for combating the Covid-19 pandemic. One such approach adopts structural controllability, a theory for controlling a network (the cell). Motivated to protect, rather than control, the cell from viral infections, we extended this theory to total structural controllability and introduced the concept of control hubs. Perturbing any control hub may render the cell uncontrollable by exogenous stimuli like viral infections, so control hubs are ideal drug targets. We developed an efficient algorithm for finding all control hubs and applied it to the largest homogenous network of human protein-protein interactions and interactions between human and SARS-Cov-2 proteins. We identified 65 druggable control hubs enriched with antiviral functions and used them to prioritize drugs into four major categories: antiviral and anti-inflammatory agents, drugs on central nerve systems, dietary supplements, and hormones that help boost immunity. The functions of these druggable control hubs and their corresponding drugs validated our new method. One of the identified drugs is Fostamatinib, which was originally approved for chronic immune thrombocytopenia and is currently in a clinical trial for treating Covid-19. As shown in a preclinical trial, Fostamatinib can lower mortality, shorten the length of ICU stay, and reduce the disease severity of hospitalized Covid-19 patients. Fostamatinib targets 10 control hubs, 9 of which are kinases playing key roles in cell proliferation, differentiation, and programmed death. One of these kinases is RIPK1 which directly interacts with the RNA-dependent RNA polymerase of SARS-CoV-2. The functions and biological processes that the druggable control hubs were involved with have not only revealed the diverse cellular responses during SARS-CoV-2 infection but provided deep insight into the therapeutic mechanisms of the drugs for Covid-19 therapy, making the new approach an explainable drug repurposing method.

**Key points**

- Extend network structural controllability to total network controllability, providing a practical method for identifying critical nodes that must be protected to maintain network controllability.
- Extend and apply the new method to identifying drug targets in human protein-protein interaction networks.
- Apply the new method to identify drugs that can be reused for Covid-19 treatment.

**Introduction**

The devastating Covid-19 pandemic caused by the severe acute respiratory syndrome coronavirus 2 (SARS-CoV-2)[1,2] wreaked global havoc on all walks of life. SARS-CoV-2 and its variants have so far infected more than 766 million people and claimed more than 6.9 million lives worldwide as reported to WHO (https://covid19.who.int; as of May 2023). The numbers are climbing despite several vaccines have been administrated in many countries. The viruses, particularly the latest omicron variants, can penetrate the vaccines and spread rapidly in densely populated areas. Therefore, it is urgent to develop effective drugs for the treatment of SARS-CoV-2 infection.

Drug discovery is notoriously costly and time-consuming[3] and the development of new drugs for Covid-19 is challenging[4,5]. To shorten the period of finding effective medicine, one approach is to reposition or repurpose the drugs that were originally developed for other diseases, a major focus of drug discovery for Covid-19[6-9]. However, the space for drug repurposing is enormous[9,10]. The most popular computational approaches for drug repurposing take the perspective of systems biology or network medicine[11-16]. Among these are eminent methods based on the well-established network structural controllability[14-16]. Following the theory of structural controllability[17,18], the cell is regarded as a network of genes/proteins that can be controlled by exogenous stimuli (e.g., viral infections or medical interventions) on a set of driver nodes (i.e., proteins) so that the cell can be driven from any state to the designated state in finite time. Structural controllability has been directly adopted to repurpose drugs for treating Covid-19[15,16]. Among a large number of driver nodes, a small number of them that were drug targets led to putative reusable drugs, and the results were validated using bioinformatic methods and data in the literature[15,16]. Structural controllability has been applied to protein-protein interaction networks[19], gene regulatory networks[20], and metabolic networks[21]. The concept of driver nodes matches well with that of cancer driver genes[22] because frequent mutations (which are viewed as stimuli) in such genes may induce tumorigenesis so the concepts of structural controllability and driver nodes have been applied to finding cancer driver genes as therapeutic targets for precision cancer treatment[14,23].

While theoretically sound, it is impractical to directly apply structural controllability to drug repurposing. The key to the structural controllability of a network is a control scheme consisting of a set of control paths and their starting nodes (driver nodes or genes) that can be used to steer the network from any arbitrary state to the designated state in a finite time. Under this theory, a drug is used as an exogenous force to change the states of the cell, hopefully, from an infectious or cancerous state to a normal state. However, the normal states of a cell are typically unknown so it remains unclear what external stimuli or drugs should be used. Moreover, the control scheme is not unique, and a control scheme typically has too many driver nodes to be practically manipulated at once to control the cell.

We pushed the envelope of the theory of structural controllability. Instead of attempting to control the cell, we aimed at protecting the cell from viral infections. Specifically, we expanded our perspective from structural controllability by a single network control scheme to a global view of total controllability over all control schemes for the network. We then introduced the concept of *control hubs*, which are nodes residing on a control path of every control scheme of the network. The control hubs are the most vulnerable spots to the structural controllability of the network; a perturbation to any control hub may render the network uncontrollable by any control scheme. Therefore, control hubs are ideal drug targets for protecting the cell from exogenous influences. Moreover, exploiting control hubs as drug targets is a more practical approach for drug repurposing because control hubs are typically an order less than driver nodes as to be shown shortly. Without computing all control schemes, which is a #P-hard problem[24] (meaning no polynomial-time or efficient algorithm is known), we developed a polynomial-time (meaning efficient) algorithm for finding all control hubs for a network. We applied our novel control hub-based drug repurposing approach to the largest homogenous human protein-protein interaction (PPI) network[25] (Table S1), along with the data of PPIs between SARS-CoV-2 and human[26,27] (Table S2) and the data of drug targets[28] (Table S3), to discover a set of existing drugs that are potentially effective for treating SARS-CoV-2 infections.

# Results

We first lay out the rationale of our novel control-hub-based method and present its major steps. We then apply it to an integrated network constructed using human and SARS-CoV-2 PPI data and the data of drugs and drug targets. We compare our new method with nine existing gene selection methods, including the structural-controllability-based driver-node method, to show its performance in finding drug targets for Covid-19. We then examine the 65 drug targets and the corresponding drugs identified by our new methods, using the data and results in the literature for validation.

## *Total network controllability for drug repurposing*

The primary concept of network structural controllability[17,18] is a control scheme for a network. It consists of control paths such that every node in the network can be reached or controlled by the head node of the control path to which the node belongs (Figure 1A). The head node is referred to as the driver or input node of the path. By exerting stimuli on the driver nodes, the network can be steered from any initial state to the designated state in a finite time. Structural controllability has been directly applied to repurposing drugs for treating Covid-19 where a small number of driver nodes targeted by drugs were used to find reusable drugs[15,16].

However, driver nodes are a double-edged sword and can also be exploited by viruses to infect the cell. Viral infections are exogenous stimuli to the cell via the interactions of viral proteins and host receptors, which can transform the cell from normal states to abnormal states to accommodate viral replication and propagation. During SARS-CoV-2 infection, the viral spike protein S engages human receptor angiotensin-converting enzyme 2 (ACE2) to enter the host cell and consequentially trigger a series of adverse signaling cascades[29].

Moreover, it is impractical to directly adopt structural controllability for controlling the cell or repurposing drugs. The control scheme is not unique (Figure 1A). An exponential number of control schemes may exist, and one control scheme may have as many as half of the nodes in the network as driver nodes. For example, one control scheme for the human PPI network[25] (Table S1) contains 4,529 driver nodes, which are 49.8% of the 9,092 nodes in the network. Determining the best or an effective control scheme is a daunting task.

In light of these serious issues underlying the approaches to controlling the cell, we resorted to protecting the cell instead. We were motivated to identify critical genes, which, when perturbed, can render the cell uncontrollable by any control scheme or external stimulus on the driver nodes. Manipulating any of such critical genes can invalidate all the control schemes, so the cell is uncontrollable by undesired stimuli. To identify such critical genes, we extended structural controllability to *total controllability* by considering all control schemes and introducing a new concept of control hubs. A *control hub* is a middle node in one of the control paths of *every* control scheme (Figure 1B). Blocking a control hub will block at least one control path of every control scheme, making the overall network uncontrollable.

Therefore, control hubs are ideal drug targets for protecting the cell from being manipulated by viral infections. If the genes that viruses act on are known, the control hubs close to these nodes can be chosen as designated drug targets to increase drug efficacy.

Since the concept of control hubs is built atop all control schemes, a technical obstacle is the potentially exponential number of control schemes for a network. Finding all control schemes using the current best method, i.e., maximum matching[30], is a computationally infeasible #P-complete problem[24]. To circumvent this difficulty, we developed an efficient, polynomial-time algorithm for finding all control hubs without computing all control schemes. The algorithm identified 1,256 control hubs in the human PPI network[25], which are 13.8% of all the 9,092 genes and 27.7% of the 4,529 driver genes for the network (Table S1).

Control hubs can act as surrogates to reusable drugs, i.e., we focus on those existing drugs that can target control hubs. While in theory, any drug-targeted control hubs can be used, the ones that are closer to exogenous stimuli (i.e., viral proteins) are preferred over the distant ones since blocking the former may prevent the spread of external influences sooner and more effectively.

*Finding drug targets for the treatment of viral infections*

We capitalized on the concepts of total controllability and control hubs and developed a drug-purposing method consisting of four major steps (Figure 1C, see Methods and Supplemental Method S3). The first is the construction of a network to integrate information on human PPI, virus PPI, drugs, and their targets. We used the largest homogenous human PPI network[25] (Table S1) and the data of PPIs between SARS-CoV-2 and human[26,27] (Table S2) and the data of drug targets[28] (Table S3). The human PPI subnetwork and the virus PPI subnetwork are linked through the PPI between human and virus proteins, and the human PPI subnetwork and drug subnetwork are connected by the drug target information. The resulting network contains 9,092 nodes (proteins) from humans, 22 nodes from SARS-CoV-2, and 2,980 nodes of drugs. The overall network is relatively tight with a total of 81,953 links.

The second step is to identify control hubs[31]. To focus on Covid-19, we left the technical details of our new method for finding control hubs to Methods and Supplemental Method S3. This control-hub finding method identified 1,256 control hubs in the network.

In the third step, to identify effective drug targets and drugs, we focused on the control hubs that were known targets of the existing drugs, which were categorically referred to as *druggable control hubs* hereafter. Among the 1,256 control hubs, 160 (12.7%) were drug targets (Figure 2A).

Druggable control hubs were not equally effective for treating SARS-CoV-2 infection. Some control hubs may directly interact with viral proteins and thus are ideal drug targets, whereas many others are far away from viral proteins in the human PPI network (Figure 1C). The closer a druggable control hub to virus proteins in the network, the more effective it should be for prohibiting viral infection.

Following this reasoning, in the fourth step, we examined the druggable control hubs in the community of proteins that were $k$ steps away from the virus proteins in the PPI network, which was referred to as the $k$-step community for convenience. A smaller $k$ is preferred; the closer a control hub is to viral proteins, the more effective it is as a drug target to block viral infections. Two sets of enrichment tests, using the z-test, were performed to identify the best $k$-step community (see Methods). The first set of tests looked for the $k$-step community that was most enriched with control hubs among all $k$-step communities for different values of $k$, and the second set of tests assessed the enrichment of drug targets among the control hubs in the community chosen in the first test. The first z-test revealed that the 2-step community was most enriched with control hubs (z-score=5.28, *p*-value=$1.3e^{-7}$, Figures 2B, S1A). It hosted 677 control hubs, among which 65 were drug targets (Table S4A). The second z-test confirmed that the 2-step community was also most enriched with druggable control hubs among all $k$-step communities (z-score=28.25, *p*-value=$1.3e^{-175}$, for $k$=2, Figures 2C, S1B).

In the last step, we assessed if our novel control-hub approach was indeed the method of choice for finding drug targets. In particular, we compared it with nine existing methods, including the driver-node-based method and eight popular node ranking methods. These included node-degree centrality, neighbor-degree centrality, betweenness centrality, load centrality, closeness centrality, and eigenvector centrality, as well as Page-Rank, and k-core[32-41]. To facilitate the comparison and better understand these methods, we compared them against a statistical model of drug targets in the 2-step community. Assuming that any protein in the 2-step community was equally likely to be a drug target, the drug-target enrichment for 677 (i.e., the number of control hubs in the 2-step community) randomly selected proteins in the community should follow an empirical normal distribution (Figure 2D). This empirical distribution was adopted as a statistical baseline model of drug-target enrichment. The enrichment of the 65 druggable control hubs in the 677 control hubs in the 2-step community substantially deviated from the baseline model (*z*-score=1.53, *p*-value=0.13; Figure 2D). Likewise, the drug-target enrichment for 677 driver nodes randomly chosen from the total of 965 driver nodes in the 2-step community should also obey an empirical normal distribution (Figure 2D). The drug-target enrichment of our control-hub method was significantly better than that of the driver-node method (*z*-score=2.82, *p*-value=0.005). The driver-node method was slightly worse than the baseline model since the mean of the former was smaller than the mean of the latter (54.07 vs 56.05; Figure 2D) and the two distributions were statistically indistinguishable ($p$-value = 0.98, $\chi^2$-test; Figure 2D). We measured the drug-target enrichments of the top 677 nodes from the eight gene-ranking methods. Unfortunately, these methods all underperformed; their z-tests against the random baseline model all resulted in negative z-scores (Figure 2D). For instance, the Page-Rank method had a z-score=-1.89 with $p$-value=0.06.

In summary, this analysis showed that our novel control-hub method can identify the largest numbers of drug targets and candidate drugs for Covid-19 treatment.

### *Control hubs as drug targets for Covid-19 treatment*

We examined the biological functions of the druggable control hubs to appreciate their role in SARS-Cov-2 infection and validate the new method using published results in the literature. Among all 160 druggable control hubs, three (RIPK1, CYB5R3, and COMT) directly interact with nonstructural proteins of SARS-CoV-2[26,27] (Figures 3A, 3B, S2; Tables 1, S4A). Remarkably, RIPK1 can bind with viral nonstructural protein nsp12[26,27], the RNA-dependent RNA polymerase (RdRp) of SARS-CoV-2[42] (Figures 3A, 3B). nsp12 not only promotes viral replication but also inhibits the host's innate immune response by suppressing the activity of interferon regulatory factor 3 (IRF3), which is key to interferon production[43]. Both CYB5R3 and COMT interact with the nsp7 protein of SARS-CoV-2 (Figures 3A, 3B), which forms a tetramer with viral nsp8[44] and functions as a cofactor of the viral RdRp, nsp12[42]. Since nsp12 and nsp7 are essential for viral transcription and replication, blocking the interactions of RIPK1 with nsp12, CYB5R3 with nsp7, and COMT with nsp7 can potentially inhibit or suppress viral replication.

RIPK1 encodes serine/threonine-protein kinase 1, plays a role in necroptosis, apoptosis, and inflammatory response, and functions as a mediator of cell death and inflammation[45]. SARS-CoV-2 infection promotes the expression of RIPK1 in the lung of Covid-19 patients and small-molecule inhibitors of RIPK1 can reduce the viral load of SARS-CoV-2 and proinflammatory cytokines in human lung organoids, indicating that the virus hijacks RIPK1-mediated immune response for its replication and propagation[46]. RIPK1 is a target of Fostamatinib (Table 1, S4A; Figure 3A), a drug that has been under intense scrutiny for treating SARS-CoV-2 infection[47-52]. Fostamatinib is an inhibitor of spleen tyrosine kinase originally approved for treating chronic immune thrombocytopenia. Fostamatinib is effective in a mouse model of acute lung injury and acute respiratory syndrome, symptoms observed in Covid-19 patients[48]. A clinical trial with a small sample of hospitalized Covid-19 patients (30 with fostamatinib versus 29 with placebo) showed that Fostamatinib can lower mortality, shorten the length of ICU stay, and reduce the disease severity of critically ill patients[49].

CYB5R3 encodes NADH-cytochrome B5 reductase 3, a flavoprotein with oxidation functions. It is targeted by three drugs (Tables 1, S4A), two of which (NADH and Flavin adenine dinucleotide) are under clinical investigation for Covid-19 treatment. NADH is an energy booster for treating chronic fatigue syndrome and improving high blood pressure and jet lag, among many other symptoms. NADH, i.e., nicotinamide adenine dinucleotide (NAD)+ hydrogen (H), is the central catalyst of cellular metabolism, a chemical naturally produced in humans and plays a role in ATP production. The SARS-CoV-2 genome does not encode enzymes for ATP generation and the virus needs to hijack host functions for viral synthesis and assembly. Therefore, NAD is regarded as a battlefield for viral infection and host immunity[53]. Indeed, coronavirus infection dysregulates the NAD metabolome, as indicated in a preclinical study[54]. Moreover, early phases 2 and 3 clinical trials showed that medication of NADH in a mixture of two metabolic activators can significantly shorten the time to complete recovery of SARS-CoV-2 infection[55].

COMT encodes catechol-O-methyltransferase that can degrade estrogens, catecholamines, and neurotransmitters such as dopamine, epinephrine, and norepinephrine. It is targeted by 14 FDA-approved drugs including Conjugated estrogens (Tables 1, S4A). Conjugated estrogens are a mixture of estrogen hormones for the treatment of hypoestrogenism-related symptoms. Estrogen has been indicated as a susceptibility factor of SARS-CoV-2 infection[56], as women are less susceptible to Covid-19[57,58] and mice with weaker estrogen receptor signaling due to respiratory coronavirus infection exhibit increased morbidity and mortality[59].

Beyond the 3 druggable control hubs that directly interact with viral proteins, 19 druggable control hubs in the 2-step community engage more than one viral protein via another protein, and four of them (SLC10A1, SLC10A6, MUC1, and TTPA) are targeted by more than one drug (Tables 1, S4A; Figure S2). The potential of these four druggable control hubs for Covid-19 treatment is discussed in Supplemental Result S1.

In short, the 65 druggable control hubs within the 2-step community were enriched with biological functions related to cell (particularly leukocyte) proliferation, cellular response to (chemical) stress, regulation of apoptotic signaling, and response to nutrient levels (Figure 3C). All these results combined revealed the essential roles that these control hubs may play in prohibiting the replication and proliferation of SARS-CoV-2. The results also revealed the essential immune-related signaling pathways that were induced by the virus and paved the way for understanding and explaining the therapeutic mechanisms of the drugs for Covid-19 treatment.

*Drugs for the treatment of SARS-CoV-2 infection*

The 65 druggable control hubs within the 2-step community were targeted by 185 existing drugs (Tables 2, S5; Figure 3D). As of June 2022, 38 were under clinical trials (https://clinicaltrials.gov/ct2/home). It is desirable to use drugs that have multiple targets to gain treatment efficacy; the potency of a drug can be estimated by the number of control hubs that it targets. Remarkably, 15 drugs target more than one control hub and 7 drugs target more than 2 druggable control hubs (Tables 2, S5).

Among the 7 drugs targeting more than 2 control hubs were Fostamatinib, NADH, and three calcium dietary supplements (Tables 2, S5). Fostamatinib is currently in a phase 3 clinical trial after a promising phase 2 trial for Covid-19 treatment[49]. Experimental and clinical data showed that Fostamatinib inhibits neutrophil extracellular traps (NETs), which entrap and eliminate pathogens during viral and bacterial infections and may cause adverse injury to surround tissues by themselves or by increasing pro-inflammatory responses[60] (Figure 3E). Activation and overreaction of innate and adaptive inflammatory responses during SARS-Cov-2 infection induce NETs which contribute to immunothrombosis in ARDS commonly seen in Covid-19 patients[47,61-63]. Moreover, coherent anti-viral therapeutic functions of Fostamatinib emerged after examining the functions of the control hubs that the drug targets (Figures 3D, 3E; Table S5). Among the 10 control hubs that Fostamatinib targets, 7 (RIPK1, CLK2, CLK3, PAK5, STK3, PKN1, and CDK4) are serine/threonine type protein kinases and two (BLK and YES1) encode Src family tyrosine kinases, all of which play essential roles in cell proliferation, cell differentiation, and programmed cell death[64]. CLK2 and CLK3 encode members of the serine/threonine type protein kinase family, and PAK5, STK3, PKN1, and CDK4 encode, respectively, one of the three members of the group II PAK family of serine/threonine kinases, serine/threonine-protein kinase 3, serine/threonine protein kinase N, and cyclin-dependent serine/threonine kinase. Plus, RIPK1 encodes receptor-interacting serine/threonine-protein kinase 1 and directly interacts with the viral RdRp nsp12 as discussed earlier. Interestingly, while not being a kinase, the remaining target COQ8A encodes a mitochondrial protein functioning in an electron-transferring membrane protein complex in the respiratory chain. Its expression is induced by the tumor suppressor p53 in response to DNA damage, and inhibition of its expression suppresses p53-induced apoptosis. Combined, the inhibitory function on NETs and kinase functions of 9 of the 10 control hubs targeted by Fostamatinib suggested it to be potent for Covid-19 treatment by acting broadly on components of autoimmune, tumor repression, and inflammatory viral response pathways (Figure 3E).

NADH targets 5 control hubs, including CYB5R3 and NDUFB7. CYB5R3 encodes NADH-cytochrome B5 reductase 3, and NDUFB7 is a subunit of the multi-subunit NADH:ubiquinone oxidoreductase. NDUFB7 functions in the mitochondrial inner membrane and has NADH dehydrogenase and oxidoreductase activities. It has been reported that the NADH level was decreased in Covid-19 patients[53] and coronavirus infection dysregulates the NAD metabolome[54], so medication of NADH plays a role in attenuating the impact of virus infection.

The three calcium dietary supplements, Calcium Citrate, Calcium Phosphate, and Calcium phosphate dihydrate, target three control hubs, including S100A13 and PEF1 which are calcium-binding proteins. A low serum calcium level has been indicated by various clinical studies as a prognostic factor of the mortality, severity, and comorbidity of SARS-CoV-2 infection[65-67]. As a side note, six vitamin E-related drugs targeting control hub TTPA (Table S5), which encodes a soluble protein that is a form of vitamin E (Tables 1, S5), have entered clinical trials for Covid-19 treatment. Combined, these results indicated that calcium, vitamin E, and many other micronutrients should be adopted as adjuvant therapy against viral infection.

In summary, the repurposed drugs fall into four major categories (Table S5), 1) antiviral and anti-inflammatory agents that are subscribed for virus infection and cancer treatment, 2) dietary supplements including NADH and Calcium that boost human immunity, 3) hormones, including conjugated estrogens, and 4) drugs acting on central nerve systems. Combined, the medicines in the first three categories help boost immunity to overcome the adverse stress and influence of viral infections.

**Discussion**

Network medicine for drug repurposing has gained popularity and momentum since the Covid-19 pandemic[11-16]. Most of these network-biology methods hinge upon the idea that important proteins can be used as surrogates for identifying medicines. However, they operate under different notions of what constitutes important proteins in biological networks. For example, proteins with high degrees of connectivity may be considered important since they supposedly affect many neighboring proteins.

Network structural controllability[17,18] has been adopted as an approach to network medicine. The idea of using driver nodes as drug targets is particularly appealing for Covid-19 drug repurposing[15,16]. However, while theoretically sound, this approach is impractical for drug repurposing as discussed earlier. Our drug-target enrichment analysis showed that such a direct application of structural controllability was no better than random selection (Figure 2D).

Our most important contributions are the extension of structural controllability to total controllability and the new perspective of protecting rather than controlling cells. In particular, we were motivated to protect the cell from any exogenous stimulus, particularly viral infections, because this is relatively easier and more effective than controlling the cell. Methodologically, by extending structural controllability to total controllability, we introduced control hubs to identify the critical spots in the cell that were important for the controllability of the cell. We used targeting drugs as external influences to make the cell uncontrollable by any viral infection. Therefore, control hubs are an effective vehicle for drug repurposing, as we demonstrated in the current study. It is not coincidental that many control hubs are also targets of existing drugs, as shown in our drug-target enrichment analysis (Figure 2D). Rather, the result revealed that proteins that have biological importance, particularly those related to immunity, resided in critical positions in the human PPI network.

To treat or prevent viral infections, control hubs in the human PPI network should be protected by blocking their interactions with viral proteins or interactions with one another which can prevent or curtail the spread of viral influence. Control hubs are thus excellent candidate drug targets for the treatment and prevention of Covid-19. The identification of such drug targets was completely data-driven and used no information on gene functions. The information on drug targets from DrugBank was brought to the analysis at a late stage of drug repurposing. Note that we used highly confident homogenous human and SARS-CoV-2 PPI data from HEK293T cells under well-controlled conditions[25-27] to avoid possible false-positive results from heterogeneous data.

Most viral proteins interacting with human proteins are nonstructural and many of them are responsible for viral transcription and replication as well as suppression of the innate and adaptive immune responses of the host (Tables S4, S7). Many druggable control hubs have immunity and antiviral functions such as regulation of apoptotic signaling, regulation of cellular response to stress, regulation of leukocyte proliferation, and regulation of cell population proliferation (Figure 3C; Tables 1, S4). Nutrient levels are another key factor that these control hubs responded to (Figure 3C; Tables 1, S4,). These results of druggable control hubs provided deep insight into the possible therapeutic mechanisms of the identified drugs for Covid-19 treatment (Figure 3E; Tables 2, S5), making our new method an explainable drug repurposing approach, which is a desirable feature for repurposing drugs[12]. For example, RIPK1 interacts with viral RdRp nsp12, and CYB5R3 and COMT interact with nsp7, a cofactor of viral RdRp (Figures 3A, 3B). RIPK1 is targeted by Fostamatinib, CYB5R3 by three drugs including NADH, and COMT by 15 drugs including Conjugated estrogens (Tables 2, 3, S4A, S5). These drugs are effective for treating Covid-19, as supported by experimental and clinical data, by blocking or suppressing the transcription or replication of SARS-Cov-2 to protect the host immunity.

One of the most important results from the new method is the identification of Fostamatinib as a Covid-19 drug, particularly suitable for hospitalized patients (Figures 3E; Tables 2, S5). This drug is currently in clinical trials for Covid-19. The identification of Fostamatinib and other drugs that are currently available for Covid-19 treatment is a strong validation of our novel method. The functions of the ten control hubs targeted by Fostamatinib explain well the mechanistic mode of action that the medicine may perform in the treatment of severely ill Covid-19 patients. It is encouraging that this data-driven result was supported by the experimental results on a mouse model of acute lung injury and acute respiratory syndrome[48] and the data of a preliminary clinical trial of critically ill patients[49]. Altogether, the biological functions and experimental data suggested that the drug functions to prevent exaggerated autoinflammatory immune responses[68,69] and alleviate the burden of cytokine storms[70,71] in severe Covid-19 cases.

A substantial number of control hubs in the 2-step community of the human PPI network are not targets of any existing drug. These control hubs, particularly the membrane proteins and those function on the NF-κB pathway (Figure S4), can be used to propose testable hypotheses for new drug development for Covid-19 therapy.

**Materials and Methods**

*An overview of the novel control-hub-based method for drug repurposing*

The new method consists of the following four major steps whose details are discussed in the subsequent subsections.

1. Construction of a biological network. In the current study, an integrated network of human PPI, virus PPI, drug targets, and drugs;
2. Identification of control hubs[31]; the algorithmic details are in Supplemental Method S3.
3. Determination of the k-step network community with nodes k steps away from the viral proteins and enriched with drug targets;
4. Assessment and validation of the new method by comparison with nine existing gene selection methods, including the structural-controllability-based driver-node method.

*Construction of a triple-layer interaction network from viruses to humans to drugs*

The central layer of the network contained the human protein-protein interaction (PPI) network that was constructed using the human Huri-Union binary protein interaction dataset[25]. This is the largest homogenous human protein interactome with data collected primarily from HEK923T cells and validated in multiple orthogonal assays. The network consists of 9,092 nodes or proteins and 64,006 interactions (Table S1).

To include the layer of viral proteins, the SARS-CoV-2 AP-MS data[26] from HEK293T cells were added. The dataset contains 332 high-confidence virus-host interactions between 27 SARS-CoV-2 proteins and 332 human proteins, which were used to link the human and virus PPI subnetworks. Since the human PPI network contains only 9,092 proteins, the final triple-layer network contains 169 interactions between 22 viral proteins and 169 human proteins (Table S2). The 3D Structural Interactome between SARS-CoV-2 and host proteins was retrieved from SARS-CoV-2-Human Interactome Browser[72].

The network was further expanded to include the layer of drugs and their human protein targets using the data from DrugBank[28]. The links between drugs and their protein targets were used to link the human PPI subnetwork and drug subnetwork. We only included drugs approved by FDA and under investigation. The drug-target interactome contains 17,780 interactions between 2,981 drugs and 2,914 target proteins (Table S3). The information on drug categories in DrugBank was used to group drugs (Table S5).

*Identification of control hubs*

A network can be controlled by exerting control signals on *driver nodes*[17,29] (Figure 1A). To analyze the controllability of a network, maximum matching from graph theory was adopted to find the minimum set of driver nodes[18]. A *maximum matching* is the maximum set of edges that do not share nodes in common[30]. The edges of a maximum matching form paths of the network, which start from *head nodes*, and along the matching edges, reach tail nodes. The head nodes of a maximum matching are taken as driver nodes and the paths are *control paths*[73] (Figure 1A), which constitute a *control scheme*. The maximum matching is not unique for most networks, neither is the control scheme (Figure 1A).

A node may occupy distinct positions – a driver, a tail, or a middle node – in control paths of different control schemes. Some nodes may always remain as middle nodes in all control schemes, and such nodes are defined as *control hub*s (Figure 1B). All control hubs can be identified in polynomial time without computing all control schemes[31]; the algorithmic details are available in our previous work[31].

*Identification of druggable control hubs within k-step from viral proteins and candidate drugs*

To find the control hubs that were reachable within no more than *k* steps from some viral proteins, a breadth-first traversal of the triple-layer PPI network was carried out. The traversal started from the viral proteins and ignored edge directions. The process terminated after all nodes at *k* steps from the beginning were visited.

All control hubs encountered in the process of the breadth-first traversal were reported. These control hubs were further checked against DrugBank[28] to identify druggable control hubs.

The best value of *k* for the *k*-step community was determined by two z-tests as described in the main text, along with the statistical significance of the two-tailed p-value. The z-tests were done using the following formulas:

$$\frac{D_F - mean(S_F)}{SD\ of\ S_F}$$

where $D_F$ is the number of nodes in the *k*-step community overlapping with druggable control hubs or control hubs, $mean(S_F)$ is the average number of druggable control hubs or control hubs overlapping with a random set of nodes of the same size as the *k*-step community, and $SD\ of\ S_F$ is the standard deviation of $S_F$ from 1,000 randomly chosen sets of nodes in the community. The details are in Supplemental Method S2.

Subsequently, we conducted a stability analysis on the druggable control hubs within the k-step communities, leveraging a random network redirection method for the Protein-Protein Interaction (PPI) network. The functionality of these proteins as druggable control hubs in the network was rigorously evaluated. Details of this analysis process can be found in Supplemental Method S4.

*Node ranking methods*

Nine popular node ranking methods were used to compare with the new control-hub-based method. These include two methods related to node degree (degree centrality[32] and average neighbor degree[33]), three related to network shortest paths (betweenness centrality[34], load centrality[35,36], and closeness centrality[37]), three related to network structures (eigenvector centrality[38,39], clustering coefficient[74], and K-core[41]), and a classical web ranking algorithm (page rank[75,76]). A detailed description of these ranking algorithms is in networkx[77] and Supplemental Method S1.

*Feature enrichment analysis*

See Supplemental Method S2 for details.

*Gene Enrichment Analysis*

To explore the biological processes in which the 65 druggable control hubs were involved, functional annotation analyses with Kyoto Encyclopedia of Genes and Genomes (KEGG) pathway annotation and Gene Ontology (GO) annotation were performed using Metascape[78]. The Go biological process terms and KEGG pathways with FDR-corrected *p*-value < 0.05 were reported.

**Declarations**

*Ethics approval and consent to participate*

Not applicable

*Consent for publication*

Not applicable

*Availability of data and materials*

The datasets used in the study are in Tables S1-S3. The datasets and software of our method are freely available on GitHub at https://github.com/network-control-lab/control-hubs.

*Competing interests*

The authors declare that they have no competing interests

*Funding*

This work was supported in part by the National Natural Science Foundation of China (grant 62176129), the United States National Institutes of Health (grant R01-GM100364), and the Hong Kong Global STEM Professorship Scheme.

*Authors' contributions*

Weixiong Zhang and Xizhe Zhang conceived and designed the research and coordinated the project. Xinru Wei and Chunyu Pan implemented the algorithm, collected and analyzed the data, and helped with writing the manuscript and preparing the figures. Weixiong Zhang analyzed the data and draft the manuscript.

*Acknowledgements*

Not applicable

**Reference**

1    Gorbalenya, A. E. *et al.* The species Severe acute respiratory syndrome-related coronavirus: classifying 2019-nCoV and naming it SARS-CoV-2. *Nature Microbiology* **5**, 536-544 (2020).
2    Chen, G. *et al.* Clinical and immunological features of severe and moderate coronavirus disease 2019. *J Clin Invest* **130**, 2620-2629 (2020).
3    Beck, A., Goetsch, L., Dumontet, C. & Corvaïa, N. Strategies and challenges for the next generation of antibody-drug conjugates. *Nat Rev Drug Discov* **16**, 315-337 (2017).
4    Abd El-Aziz, T. M. & Stockand, J. D. Recent progress and challenges in drug development against COVID-19 coronavirus (SARS-CoV-2)-an update on the status. *Infection Genetics and Evolution* **83** (2020).
5    Kim, S. COVID-19 Drug Development. *J Microbiol Biotechnol* **32**, 1-5 (2022).
6    Riva, L. *et al.* Discovery of SARS-CoV-2 antiviral drugs through large-scale compound repurposing. *Nature* **586**, 113-+ (2020).
7    Dotolo, S., Marabotti, A., Facchiano, A. & Tagliaferri, R. A review on drug repurposing applicable to COVID-19. *Brief Bioinform* **22**, 726-741 (2021).

8   Chakraborty, C., Sharma, A. R., Bhattacharya, M., Agoramoorthy, G. & Lee, S. S. The Drug Repurposing for COVID-19 Clinical Trials Provide Very Effective Therapeutic Combinations: Lessons Learned From Major Clinical Studies. *Front Pharmacol* **12**, 704205 (2021).
9   Ng, Y. L., Salim, C. K. & Chu, J. J. H. Drug repurposing for COVID-19: Approaches, challenges and promising candidates. *Pharmacol Ther* **228**, 107930 (2021).
10  Pushpakom, S. *et al.* Drug repurposing: progress, challenges and recommendations. *Nat Rev Drug Discov* **18**, 41-58 (2019).
11  Sadegh, S. *et al.* Exploring the SARS-CoV-2 virus-host-drug interactome for drug repurposing. *Nat Commun* **11**, 3518 (2020).
12  Jimenez-Luna, J., Grisoni, F. & Schneider, G. Drug discovery with explainable artificial intelligence. *Nature Machine Intelligence* **2**, 573-584 (2020).
13  Morselli Gysi, D. *et al.* Network medicine framework for identifying drug-repurposing opportunities for COVID-19. *Proc Natl Acad Sci U S A* **118** (2021).
14  Guo, W. F., Zhang, S. W., Zeng, T., Akutsu, T. & Chen, L. Network control principles for identifying personalized driver genes in cancer. *Briefings in bioinformatics* **21**, 1641-1662 (2020).
15  Ackerman, E. E. & Shoemaker, J. E. Network Controllability-Based Prioritization of Candidates for SARS-CoV-2 Drug Repositioning. *Viruses* **12** (2020).
16  Siminea, N. *et al.* Network analytics for drug repurposing in COVID-19. *Brief Bioinform* **23** (2022).
17  Lin, C.-T. Structural controllability. *IEEE Transactions on Automatic Control* **19**, 201-208 (1974).
18  Liu, Y. Y., Slotine, J. J. & Barabási, A. L. Controllability of complex networks. *Nature* **473**, 167-173 (2011).
19  Kanhaiya, K., Czeizler, E., Gratie, C. & Petre, I. Controlling Directed Protein Interaction Networks in Cancer. *Sci Rep* **7**, 10327 (2017).
20  Qian, X., Ivanov, I., Ghaffari, N. & Dougherty, E. R. Intervention in gene regulatory networks via greedy control policies based on long-run behavior. *BMC Syst Biol* **3**, 61 (2009).
21  Asgari, Y., Salehzadeh-Yazdi, A., Schreiber, F. & Masoudi-Nejad, A. Controllability in cancer metabolic networks according to drug targets as driver nodes. *PLoS One* **8**, e79397 (2013).
22  Bailey, M. H. *et al.* Comprehensive Characterization of Cancer Driver Genes and Mutations. *Cell* **173**, 371-385.e318 (2018).
23  Guo, W. F. *et al.* Network controllability-based algorithm to target personalized driver genes for discovering combinatorial drugs of individual patients. *Nucleic Acids Res* **49**, e37 (2021).
24  Valiant, L. G. The complexity of computing the permanent. *Theoretical computer science* **8**, 189-201 (1979).
25  Luck, K. *et al.* A reference map of the human binary protein interactome. *Nature* **580**, 402-408 (2020).
26  Gordon, D. E. *et al.* A SARS-CoV-2 protein interaction map reveals targets for drug repurposing. *Nature* **583**, 459-468 (2020).
27  Gordon, D. E. *et al.* Comparative host-coronavirus protein interaction networks reveal pan-viral disease mechanisms. *Science* **370** (2020).
28  Wishart, D. S. *et al.* DrugBank: a knowledgebase for drugs, drug actions and drug targets. *Nucleic Acids Res* **36**, D901-906 (2008).
29  Hoffmann, M. *et al.* SARS-CoV-2 Cell Entry Depends on ACE2 and TMPRSS2 and Is Blocked by a Clinically Proven Protease Inhibitor. *Cell* **181**, 271-280.e278 (2020).
30  Hopcroft, J. E. & Karp, R. M. An n^5/2 algorithm for maximum matchings in bipartite graphs. *SIAM Journal on computing* **2**, 225-231 (1973).
31  Zhang, X., Pan, C. & Zhang, W. Control hubs of complex networks and a polynomial-time identification algorithm. arXiv:2206.01188 (2022). <https://ui.adsabs.harvard.edu/abs/2022arXiv220601188Z>.
32  Borgatti, S. P. & Halgin, D. S. Analyzing affiliation networks. *The Sage handbook of social network analysis* **1**, 417-433 (2011).
33  Barrat, A., Barthelemy, M., Pastor-Satorras, R. & Vespignani, A. The architecture of complex weighted networks. *Proceedings of the national academy of sciences* **101**, 3747-3752 (2004).

34  Brandes, U. On variants of shortest-path betweenness centrality and their generic computation. *Social networks* **30**, 136-145 (2008).
35  Newman, M. E. Scientific collaboration networks. II. Shortest paths, weighted networks, and centrality. *Physical review E* **64**, 016132 (2001).
36  Goh, K.-I., Kahng, B. & Kim, D. Universal behavior of load distribution in scale-free networks. *Physical review letters* **87**, 278701 (2001).
37  Freeman, L. Centrality in networks: I. conceptual clarifications. social networks. *Social Network* (1979).
38  Bonacich, P. Power and centrality: A family of measures. *American journal of sociology* **92**, 1170-1182 (1987).
39  Brede, M. Networks—An Introduction. Mark EJ Newman.(2010, Oxford University Press.) $65.38,£ 35.96 (hardcover), 772 pages. ISBN-978-0-19-920665-0.  (MIT Press One Rogers Street, Cambridge, MA 02142-1209, USA journals-info …, 2012).
40  Zhang, J.-X., Chen, D.-B., Dong, Q. & Zhao, Z.-D. J. S. r. Identifying a set of influential spreaders in complex networks.  **6**, 1-10 (2016).
41  Batagelj, V. & Zaversnik, M. An O (m) algorithm for cores decomposition of networks. *arXiv preprint cs/0310049* (2003).
42  Gao, Y. *et al.* Structure of the RNA-dependent RNA polymerase from COVID-19 virus. *Science* **368**, 779-782 (2020).
43  Wang, W. *et al.* SARS-CoV-2 nsp12 attenuates type I interferon production by inhibiting IRF3 nuclear translocation. *Cell Mol Immunol* **18**, 945-953 (2021).
44  Zhang, C., Li, L., He, J., Chen, C. & Su, D. Nonstructural protein 7 and 8 complexes of SARS-CoV-2. *Protein Sci* **30**, 873-881 (2021).
45  Mifflin, L., Ofengeim, D. & Yuan, J. Receptor-interacting protein kinase 1 (RIPK1) as a therapeutic target. *Nat Rev Drug Discov* **19**, 553-571 (2020).
46  Xu, G. *et al.* SARS-CoV-2 promotes RIPK1 activation to facilitate viral propagation. *Cell Res* **31**, 1230-1243 (2021).
47  Strich, J. R. *et al.* Fostamatinib Inhibits Neutrophils Extracellular Traps Induced by COVID-19 Patient Plasma: A Potential Therapeutic. *J Infect Dis* **223**, 981-984 (2021).
48  Kost-Alimova, M. *et al.* A High-Content Screen for Mucin-1-Reducing Compounds Identifies Fostamatinib as a Candidate for Rapid Repurposing for Acute Lung Injury. *Cell Rep Med* **1**, 100137 (2020).
49  Strich, J. R. *et al.* Fostamatinib for the treatment of hospitalized adults with COVD-19 A randomized trial. *Clin Infect Dis* (2021).
50  Hoepel, W. *et al.* High titers and low fucosylation of early human anti-SARS-CoV-2 IgG promote inflammation by alveolar macrophages. *Sci Transl Med* **13** (2021).
51  Apostolidis, S. A. *et al.* Signaling through FcγRIIA and the C5a-C5aR pathway mediates platelet hyperactivation in COVID-19. *bioRxiv* (2021).
52  Rivero-García, I., Castresana-Aguirre, M., Guglielmo, L., Guala, D. & Sonnhammer, E. L. L. Drug repurposing improves disease targeting 11-fold and can be augmented by network module targeting, applied to COVID-19. *Sci Rep* **11**, 20687 (2021).
53  Brenner, C. Viral infection as an NAD(+) battlefield. *Nat Metab* **4**, 2-3 (2022).
54  Heer, C. D. *et al.* Coronavirus infection and PARP expression dysregulate the NAD metabolome: An actionable component of innate immunity. *J Biol Chem* **295**, 17986-17996 (2020).
55  Altay, O. *et al.* Combined Metabolic Activators Accelerates Recovery in Mild-to-Moderate COVID-19. *Advanced Science* **8** (2021).
56  Brandi, M. L. Are sex hormones promising candidates to explain sex disparities in the COVID-19 pandemic? *Rev Endocr Metab Disord* **23**, 171-183 (2022).
57  Chen, N. *et al.* Epidemiological and clinical characteristics of 99 cases of 2019 novel coronavirus pneumonia in Wuhan, China: a descriptive study. *Lancet* **395**, 507-513 (2020).

58  Guan, W. J. *et al.* Clinical Characteristics of Coronavirus Disease 2019 in China. *N Engl J Med* **382**, 1708-1720 (2020).
59  Channappanavar, R. *et al.* Sex-Based Differences in Susceptibility to Severe Acute Respiratory Syndrome Coronavirus Infection. *J Immunol* **198**, 4046-4053 (2017).
60  Mutua, V. & Gershwin, L. J. A Review of Neutrophil Extracellular Traps (NETs) in Disease: Potential Anti-NETs Therapeutics. *Clin Rev Allergy Immunol* **61**, 194-211 (2021).
61  Middleton, E. A. *et al.* Neutrophil extracellular traps contribute to immunothrombosis in COVID-19 acute respiratory distress syndrome. *Blood* **136**, 1169-1179 (2020).
62  Bautista-Becerril, B. *et al.* Immunothrombosis in COVID-19: Implications of Neutrophil Extracellular Traps. *Biomolecules* **11** (2021).
63  Szturmowicz, M. & Demkow, U. Neutrophil Extracellular Traps (NETs) in Severe SARS-CoV-2 Lung Disease. *Int J Mol Sci* **22** (2021).
64  Capra, M. *et al.* Frequent alterations in the expression of serine/threonine kinases in human cancers. *Cancer Res* **66**, 8147-8154 (2006).
65  Torres, B. *et al.* Impact of low serum calcium at hospital admission on SARS-CoV-2 infection outcome. *Int J Infect Dis* **104**, 164-168 (2021).
66  Alemzadeh, E., Alemzadeh, E., Ziaee, M., Abedi, A. & Salehiniya, H. The effect of low serum calcium level on the severity and mortality of Covid patients: A systematic review and meta-analysis. *Immun Inflamm Dis* **9**, 1219-1228 (2021).
67  Pechlivanidou, E. *et al.* The prognostic role of micronutrient status and supplements in COVID-19 outcomes: A systematic review. *Food Chem Toxicol* **162**, 112901 (2022).
68  Tan, L. Y., Komarasamy, T. V. & Rmt Balasubramaniam, V. Hyperinflammatory Immune Response and COVID-19: A Double Edged Sword. *Front Immunol* **12**, 742941 (2021).
69  Rodríguez, Y. *et al.* Autoinflammatory and autoimmune conditions at the crossroad of COVID-19. *J Autoimmun* **114**, 102506 (2020).
70  Hu, B., Huang, S. & Yin, L. The cytokine storm and COVID-19. *J Med Virol* **93**, 250-256 (2021).
71  Jiang, Y. *et al.* Cytokine storm in COVID-19: from viral infection to immune responses, diagnosis and therapy. *Int J Biol Sci* **18**, 459-472 (2022).
72  Wierbowski, S. D. *et al.* A 3D Structural Interactome to Explore the Impact of Evolutionary Divergence, Population Variation, and Small-molecule Drugs on SARS-CoV-2-Human Protein-Protein Interactions. *bioRxiv*, 2020.2010.2013.308676 (2020).
73  Ruths, J. & Ruths, D. Control profiles of complex networks. *Science* **343**, 1373-1376 (2014).
74  Saramäki, J., Kivelä, M., Onnela, J.-P., Kaski, K. & Kertesz, J. Generalizations of the clustering coefficient to weighted complex networks. *Physical Review E* **75**, 027105 (2007).
75  Langville, A. N. & Meyer, C. D. A survey of eigenvector methods for web information retrieval. *SIAM review* **47**, 135-161 (2005).
76  Page, L., Brin, S., Motwani, R. & Winograd, T. The PageRank citation ranking: Bringing order to the web.  (Stanford InfoLab, 1999).
77  Hagberg, A., Swart, P. & S Chult, D. Exploring network structure, dynamics, and function using NetworkX. (Los Alamos National Lab.(LANL), Los Alamos, NM (United States), 2008).
78  Zhou, Y. *et al.* Metascape provides a biologist-oriented resource for the analysis of systems-level datasets. *Nature communications* **10**, 1-10 (2019).

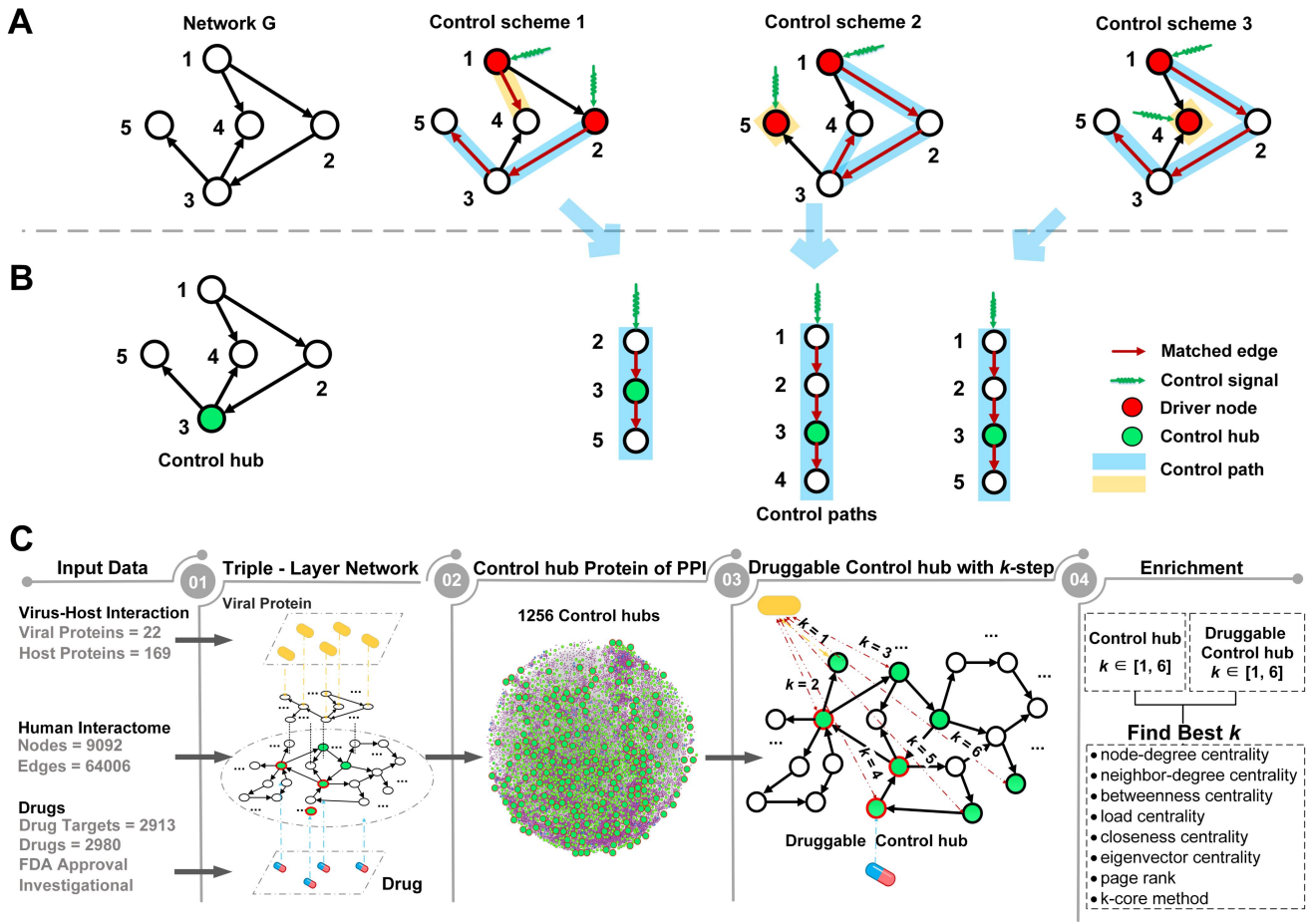

**Figure 1** | Control schemes and control hubs of a small network *G*, and a new control-hub-based approach for drug repurposing. **A)** Three distinct control schemes are identified by the maximum matching of *G*. Starting from a driver node (in red), a control path follows matched edges (in red). All control paths form a control scheme for *G*, and *G* has three control schemes. **B)** *G* has one control hub node (in green), which appears in the middle of a control path of each of the three control schemes. **C)** The study design and the framework of a new control-hub-based approach. A triple-layer network, connecting the viral and human proteins as well as drugs and human protein targets. The study focused on the network community of proteins that were no more than 2 steps away from viral proteins (i.e., the 2-step community) and the 65 druggable control hubs within the community. The enrichment of druggable control hubs within the 2-step community was assessed against that of several gene ranking methods (see main text).

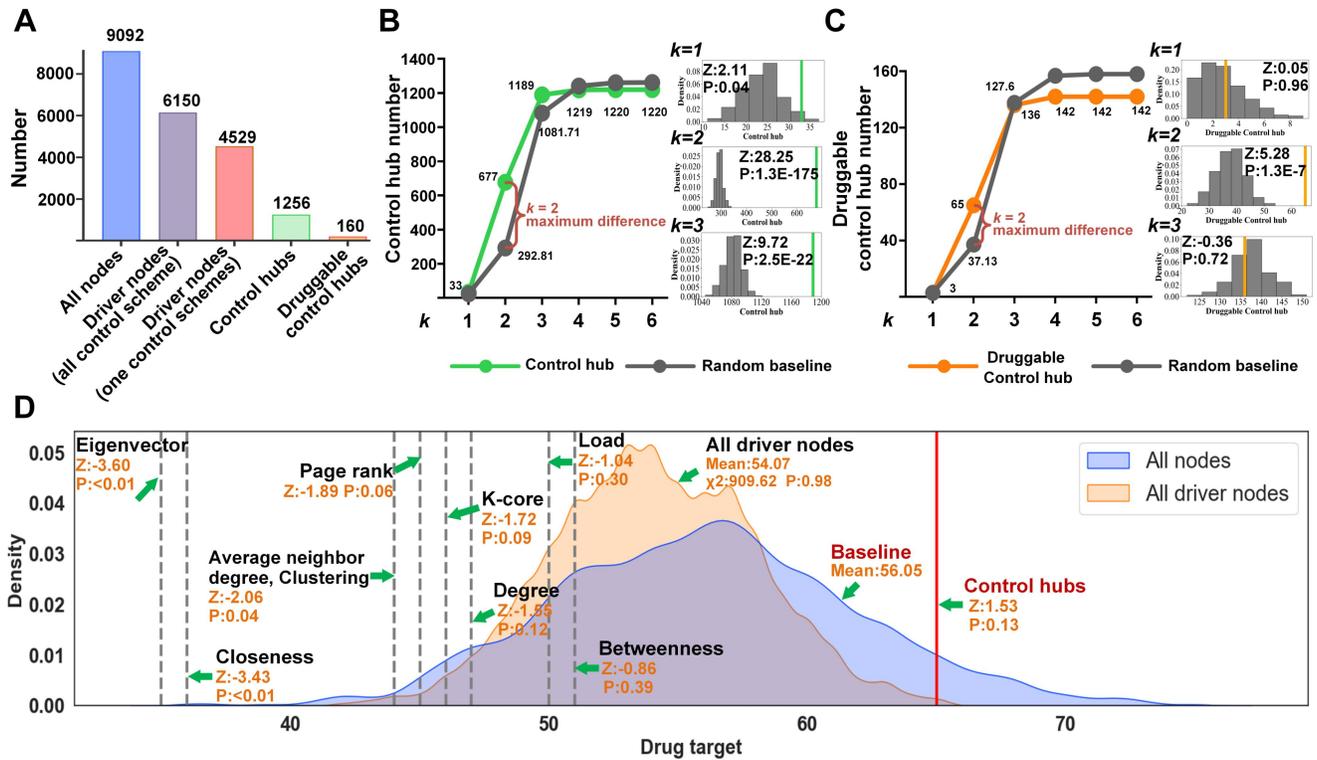

**Figure 2** | Comparison of druggable targets from different methods. **A)** Distributions of the driver nodes, control hubs, and druggable control hubs in the whole human PPI network. **B)** Determining that the 2-step community was most enriched with control hubs (the vertical axis) among all $k$-step communities of proteins with different $k$ steps away from viral proteins (the horizontal axis). Statistical analysis was adopted to compare the number of control hubs (in green) within $k$-step communities against random empirical distributions (i.e., the baseline in grey). The three smaller figures on the side show random empirical distributions for $k$=1, $k$=2, and $k$=3. Included in the small figures are the values of enrichment of druggable targets (vertical green lines) by the new control-hub method. A $z$-test analysis showed that the highest increment of control hubs from the baseline occurred at $k$=2. **C)** The 2-step community was also enriched with druggable control hubs (the vertical axis). The same statistical analysis as in B) was performed. **D)** Comparison of drug-target enrichment of the new method and that of the driver-node method and other eight node-ranking methods in the 2-step community.

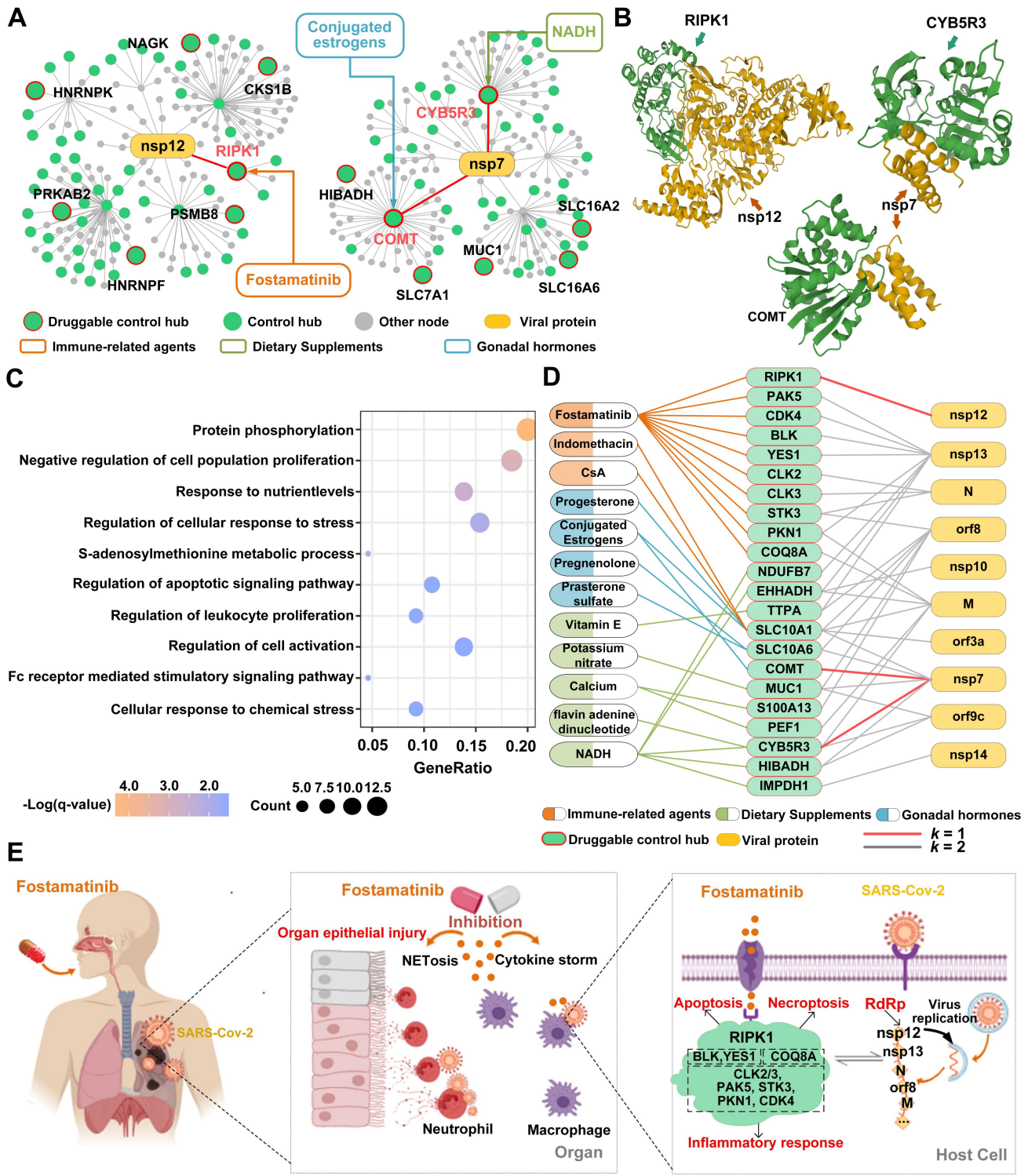

**Figure 3 | Potential therapeutic mechanisms of some druggable control hubs and some selected drugs for treatment and/or prevention of Covid-19. A)** Network topologies of two SARS-Cov-2 proteins (nsp12 and nsp7 that are responsible for viral transcription and replication) and three human proteins (RIPK1, COMT, and CYB5R3) that directly interact with nsp12 and nsp7. **B)** The binding structures of two SARS-Cov-2 proteins (nsp12 and nsp7) and three druggable control hubs (RIPK1, COMT, and CYB5R3). **C)** The biological-process enrichment of the 65 druggable control hubs within the 2-step community, revealing their collective functions

during viral infection. GeneRatio is the ratio between the number of observed proteins with a specific Go term and the total number of proteins of interest. **D)** The interactions among SARS-Cov-2 proteins, key druggable control hubs, and drugs in three categories. Drugs are grouped based on their functions, marked in color. The drugs in orange correspond to immune-related agents, such as antineoplastic or Immunomodulating agents, in green are dietary supplements, such as Vitamins and Calcium, and in blue are gonadal hormones. **E)** The potential therapeutic mechanisms of Fostamatinib for treating Covid-19. It reduces excessive immune and autoinflammatory responses by targeting 10 control hubs, 9 of which are protein kinases and one on the p53 pathway.

**Table 1** | Twenty-eight of the 65 druggable control hubs (in Table S4) within no more than two steps away from SARS-Cov-2 proteins in the triple-layer PPI network. Shown are the druggable control hub (the **Host Protein** column), engaging **Viral Protein** (and the [**Distance**] between the host and viral proteins), **Host Protein Function**, and **Targeting Drugs** (and the **Total number** of drugs targeting the protein). Drugs are grouped based on their function categories marked in color. Drugs in orange correspond to immune-related agents, in green are dietary supplements, and in blue are gonadal hormones. The seven druggable control hubs discussed in the text are marked in grey background. The rest 21 druggable control hubs are targeted by at least two drugs.

| Host Protein | Viral Protein [Distance] | Host Protein Function | Targeting Drugs (Total number) |
|---|---|---|---|
| RIPK1 | nsp12[1] | Inflammation, cell death, pathogen recognition | Fostamatinib |
| CYB5R3 | nsp7[1], orf9c[2], M[2] | Cholesterol biosynthetic process. | NADH, Flavin adenine dinucleotide, Copper |
| COMT | nsp7[1] | Neurotransmitter catabolic process | Conjugated estrogens, Diethylstilbestrol, Tolcapone (15) |
| SLC10A1 | nsp7[2], orf8[2], orf3a[2], M[2] | Bile acid and bile salt transport | Conjugated estrogens, Progesterone, Indomethacin, Cyclosporine A (CsA) (18) |
| SLC10A6 | nsp7[2], orf8[2] | Transmembrane transport | Pregnenolone, Prasterone sulfate |
| MUC1 | nsp7[2], orf9c[2], orf8[2] | Forming protective mucous barriers on epithelial surfaces | Fostamatinib, Potassium nitrate, TG4010 |
| TTPA | orf8[2], nsp13[2] | Vitamin E metabolic process | Vitamin E supplements (6) |
| OPRM1 | orf8[2] | A class of opioid receptors | Tramadol, Morphine, Codeine (47) |
| TSPO | orf8[2] | Steroid hormone synthesis, immune response | Lorazepam, Temazepam, Alprazolam (12) |
| GLUL | orf8[2] | Ammonia and glutamate detoxification, acid-base homeostasis | Pegvisomant, L-Glutamine, Methionine (8) |
| IMPDH1 | nsp14[2] | Regulate cell growth | NADH, Ribavirin, Mycophenolate mofetil (7) |
| S100A13 | orf8[2] | Cell cycle progression and differentiation | Calcium (7) |
| RARA | orf8[2] | Regulation of differentiation of clock genes. | Adapalene, Acitretin, Alitretinoin (7) |
| MAPK1 | nsp13[2] | Differentiation, transcription regulation | Isoprenaline, Arsenic trioxide (6) |
| SLC16A2 | nsp7[2] | Transporter of thyroid hormone | Pyruvic acid, Tyrosine, L-Leucine (6) |
| CDK4 | nsp13[2] | Cell cycle progression | Fostamatinib, Palbociclib, Ribociclib, Abemaciclib |
| HDAC4 | nsp13[2] | Transcriptional regulation, cell cycle progression | Zinc, Romidepsin (4) |
| PEF1 | M[2] | Response to calcium ion | Calcium (3) |
| CATSPER1 | M[2] | Calcium ion transport | Calcium (3) |
| GNMT | orf8[2] | Methionine metabolic process | Ademetionine, Glycine, Citric acid |
| PCYT1A | orf8[2] | Regulation of phosphatidylcholine biosynthesis. | Choline, Lamivudine, Choline salicylate |
| SLC7A1 | nsp7[2] | Involved in amino acid transport. | L-Lysine, L-Arginine, Ornithine |
| YES1 | nsp13[2] | Innate immune response; transmembrane receptor protein | Fostamatinib, Dasatinib |
| ASPH | orf8[2] | Calcium ion transmembrane transport | Aspartic acid, Succinic acid |
| MAT2A | nsp9[2] | Catalyzes the production of AdoMet | Ademetionine, Methionine |
| PCNA | nsp15[2] | Positive regulation of DNA repair | Liothyronine, Acetylsalicylic acid |
| SRR | nsp15[2] | Catalyzes the synthesis of D-serine from L-serine. | Pyridoxal phosphate, Serine |
| SULT2B1 | nsp8[2] | Catalyze sulfate conjugation of many hormones, and drugs. | Prasterone, Pregnenolone |

**Table 2** | Drugs for the treatment and/or prevention of Covid-19. Fifteen candidate drugs target more than one druggable control hub, among which two belong to immune-related agents (in color), nine are dietary supplements (in green) and two are gonadal hormones (in blue). The drugs with "*" are under clinical trial for the treatment of Covid-19. Detailed information is available in Table S5.

| Drug | Control hub Number | Control hub |
|---|---|---|
| Fostamatinib* | 10 | COQ8A, CLK3, CLK2, YES1, BLK, PAK5, STK3, RIPK1, PKN1, CDK4 |
| NADH* | 5 | CYB5R3, EHHADH, HIBADH, NDUFB7, IMPDH1 |
| Calcium phosphate dihydrate | 3 | S100A13, PEF1, CATSPER1 |
| Calcium Citrate | 3 | S100A13, PEF1, CATSPER1 |
| Calcium Phosphate | 3 | S100A13, PEF1, CATSPER1 |
| Conjugated estrogens* | 2 | SLC10A1, COMT |
| Progesterone* | 2 | SLC10A1, SULT2B1 |
| Phenethyl Isothiocyanate | 5 | HNRNPF, TPM1, |
| Ademetionine | 3 | COMT, GNMT, MAT2A |
| Copper | 2 | AHCY, CYB5R3 |
| Aspartic acid | 2 | ASPH, ACY3 |
| Methionine | 2 | GLUL, MAT2A |
| Citric acid | 2 | HGS, GNMT |
| Liothyronine | 2 | PCNA, SLC10A1 |
| Liotrix | 2 | SLC10A1, SLC16A2 |